\documentclass[conference,a4paper]{IEEEtran}
\IEEEoverridecommandlockouts

\usepackage{amsmath,amssymb,amsfonts}
\usepackage{graphicx}
\usepackage{textcomp}
\usepackage[english]{babel}
\usepackage{xcolor}
\usepackage{adjustbox}
\usepackage[sorting=none]{biblatex}
\usepackage{pdfpages}
\usepackage{caption}
\usepackage{subcaption}
\usepackage{diagbox}
\usepackage{makecell}
\usepackage{steinmetz}
\usepackage[unicode,pdfmenubar,linktoc=all,hidelinks,bookmarks]{hyperref}
\usepackage[nameinlink,compress,sort]{cleveref}
\graphicspath{{figures/}}

\usepackage{orcidlink}

\crefname{figure}{Fig.}{Fig.}
\Crefname{figure}{Fig.}{Fig.}

\def\BibTeX{{\rm B\kern-.05em{\sc i\kern-.025em b}\kern-.08em
    T\kern-.1667em\lower.7ex\hbox{E}\kern-.125emX}}

\begin{document}

\title{Ancillary Services Provision by Cross-Voltage-Level Power Flow Control using Flexibility Regions
\thanks{The authors gratefully acknowledge funding by the German Federal Ministry of Education and Research (BMBF) within the Kopernikus project ENSURE ‘New \textbf{EN}ergy grid \textbf{S}truct\textbf{UR}es for the German \textbf{E}nergiewende’.}}

\author{\IEEEauthorblockN{Christian Holger Nerowski \orcidlink{0009-0001-1297-9936}, Zongjun Li, Christian Rehtanz \orcidlink{0000-0002-8134-6841}}
\IEEEauthorblockA{
    \textit{Institute of Energy Systems, Energy Efficiency and Energy Economics} \\
    \textit{TU Dortmund University - Dortmund, Germany} \\
    E-Mail: christian.nerowski@tu-dortmund.de}
}

\maketitle

\begin{abstract}
    The large-scale integration of distributed renewable energy sources into the electricity grid requires the investigation of new methods to ensure stability.
    For example, Active Distribution Networks~(ADNs) can be used at (sub-)~transmission levels for emergency operation, provided robust and efficient control is available.
    This paper investigates the use of Feasible Operating Regions~(FORs) and Flexibility Regions~(FRs) for Cross-Voltage-Level Power Flow Control~(CPFC).
    The enhancement of network stability due to the provision of ancillary services is illustrated, as is the need for strengthened cooperation between Transmission~(TSOs) and Distribution System Operators~(DSOs).
    Optimal power flow methods are considered, focusing on computational advances through PieceWise~Linearization~(PWL) and convex relaxation techniques aiming to speed up runtime while keeping high accuracy.
    To illustrate the algorithms' benefits and drawbacks, they are analyzed using exemplary medium voltage grids.
\end{abstract}


\section{Introduction}\label{Introduction}

With the energy transition, the restructuring of the electrical energy system continues to progress.
The shift from centralized to decentralized units in electricity generation  will lead to fewer controllable assets in transmission grids, but in turn to more at distribution levels~\cite{ENTSOE.2020}.
In future, the installed capacity of renewable energies will exceed peak load demand in Europe notably~\cite{ENTSOE.2022}.
For this reason, situations will become frequent when decentralized controllable assets have to ensure stability, particularly of voltages and frequency.
Therefore, network operation must adapt to these new conditions~\cite{BMWK2.2023}.

To counteract the lack of control options in transmission grids, one approach is to control renewable energy sources in clusters.
Such groups of assets can be allocated inside of Active Distribution Networks~(ADNs)~\cite{DMG.2018}.
By controlling inverter-based resources and on-load-tap changers, ADNs become able to provide aforementioned support to higher levels.
In principle, these ADNs depict radial grids whose operating state is customizable using their renewable energy sources.
They rely heavily on information and communication technology to guide their assets' behavior.
Said features enable the support of superimposed levels through the Point of Common Coupling~(PCC) as discussed in numerous publications.
Some approaches such as~\cite{Valverde.2013} are based on Model Predictive Control or Optimal Power Flow~(OPF)~\cite{SaintPierre.2017}.
Others use PI-based concepts such as~\cite{Robitzky.2017}, characterized by Cross-Voltage-Level Power Flow Control~(CPFC) enabling real-time management.

Due to increased prevalence of fast-acting inverter-based resources, providing time-critical services will become possible.
In principle, an emergency requires the electrical system to be kept in balance within seconds to minutes.
The response time of a CPFC must therefore comply with hard limits, as failure is not tolerated.
As discussed in~\cite{BMWK2.2023}, reaction periods will change significantly in the future, requiring rapid actions.
According to~\cite{Zwartscholten.2022}, this results in two possible CPFC-applications:

\begin{itemize}
    \item Frequency control (by providing active power control through subordinate ADNs due to fast-acting assets)
    \item Voltage regulation for transmission grids (by providing reactive power regulation through subordinate ADNs)
\end{itemize}

This paper is structured as follows: In~\Cref{Modeling Flexibility Providing Units}, we define what characterizes Flexibility Providing Units~(FPUs), Feasible Operating Regions~(FORs) and Flexibility Regions~(FRs).
Henceforth, a conceptual framework for TSO/DSO interaction is presented in~\Cref{Determination of FORs and FRs}, as well as the ac-OPF for benchmarking.
In~\Cref{Accelerated calculation methods}, various accelerated computation methods are described and used in~\Cref{Test System Results} to present test case results.
Finally, conclusions are drawn in~\Cref{Conclusion and Outlook}.

\IEEEpubidadjcol

\section{Modeling Flexibility Providing Units}\label{Modeling Flexibility Providing Units}

\begin{figure}
    \centering
    \includegraphics[width=0.35\textwidth]{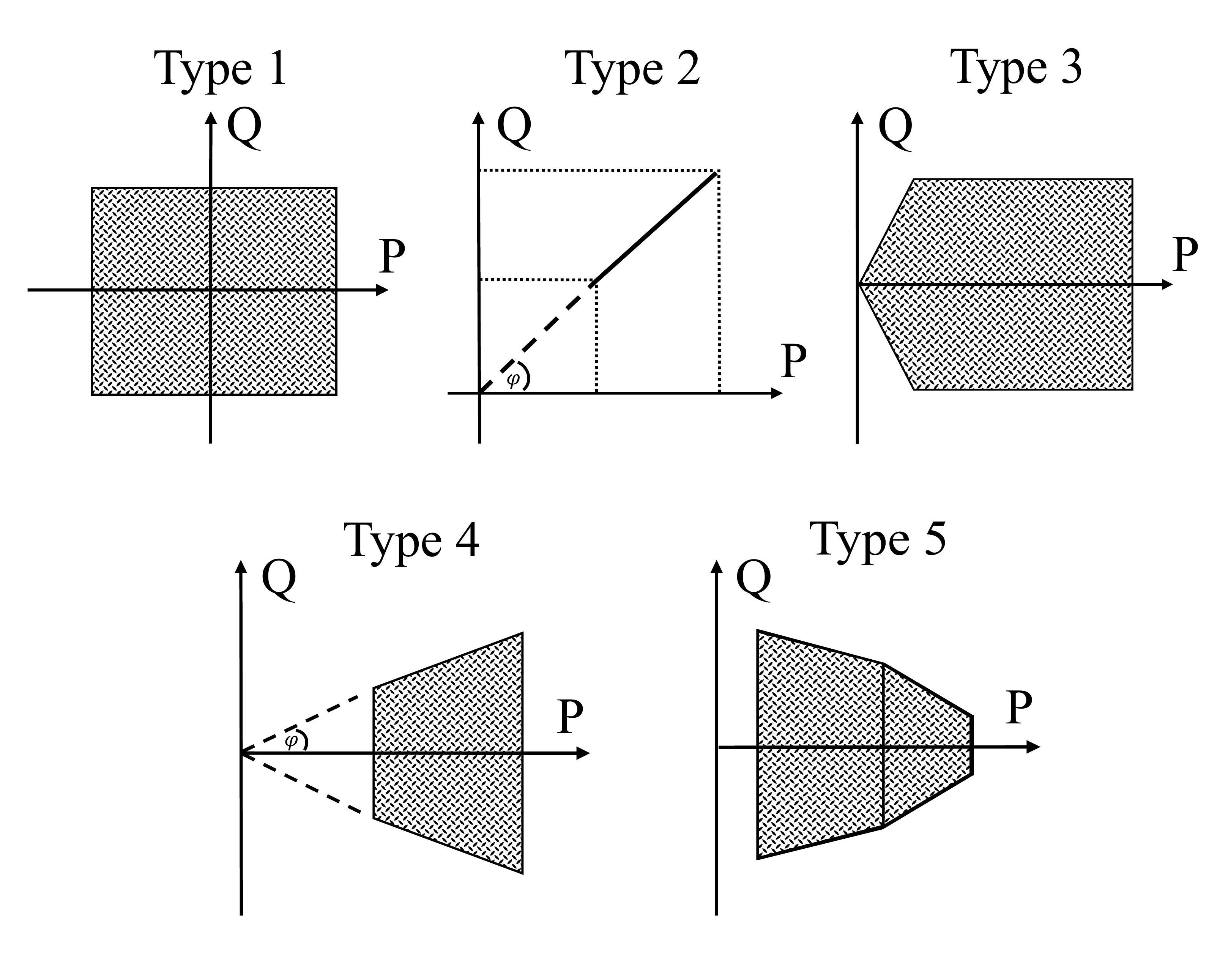}
    \caption{Classification of Flexibility Providing Units according to~\cite{Contreras.2018b}}
    \label{FPUs divided by respective FORs}
\end{figure}

As indicated, each asset needs to be managed in real-time to enable frequency control and voltage regulation.
These are often referred to as FPUs if they are geographically distributed and remotely controllable via external signals.
As described in~\cite{Contreras.2018}, there is a wide range of FPUs with various properties.
Depending on the ability to achieve certain active~($P$) and reactive~($Q$) power combinations, five types are differentiated by their FORs.
Here, any FOR results from asset's operating limits, which in turn are based on physical and technical constraints.
In short, FORs describe theoretically achievable operational states of individual assets or entire network areas, reachable only if no external time-dependent factors restrict them.
However, if these non-influenceable given factors, e.g. solar irradiation or wind strength, come into play, they reduce available $PQ$-spaces resulting in FRs.
FRs are thereby subsets of FORs.
As proposed in~\cite{Contreras.2018b}, we assume FPUs comprehend batteries~(1), controllable loads~(1) with constant $\cos(\phi)$ (2), static synchronous compensators~(1) and synchronous generators~(5) as shown in \Cref{FPUs divided by respective FORs}. Moreover, photovoltaic systems~(4) and wind turbines~(3,4) are considered to be FPUs.

\begin{figure}[b]
    \centering
    \includegraphics[width=0.4\textwidth]{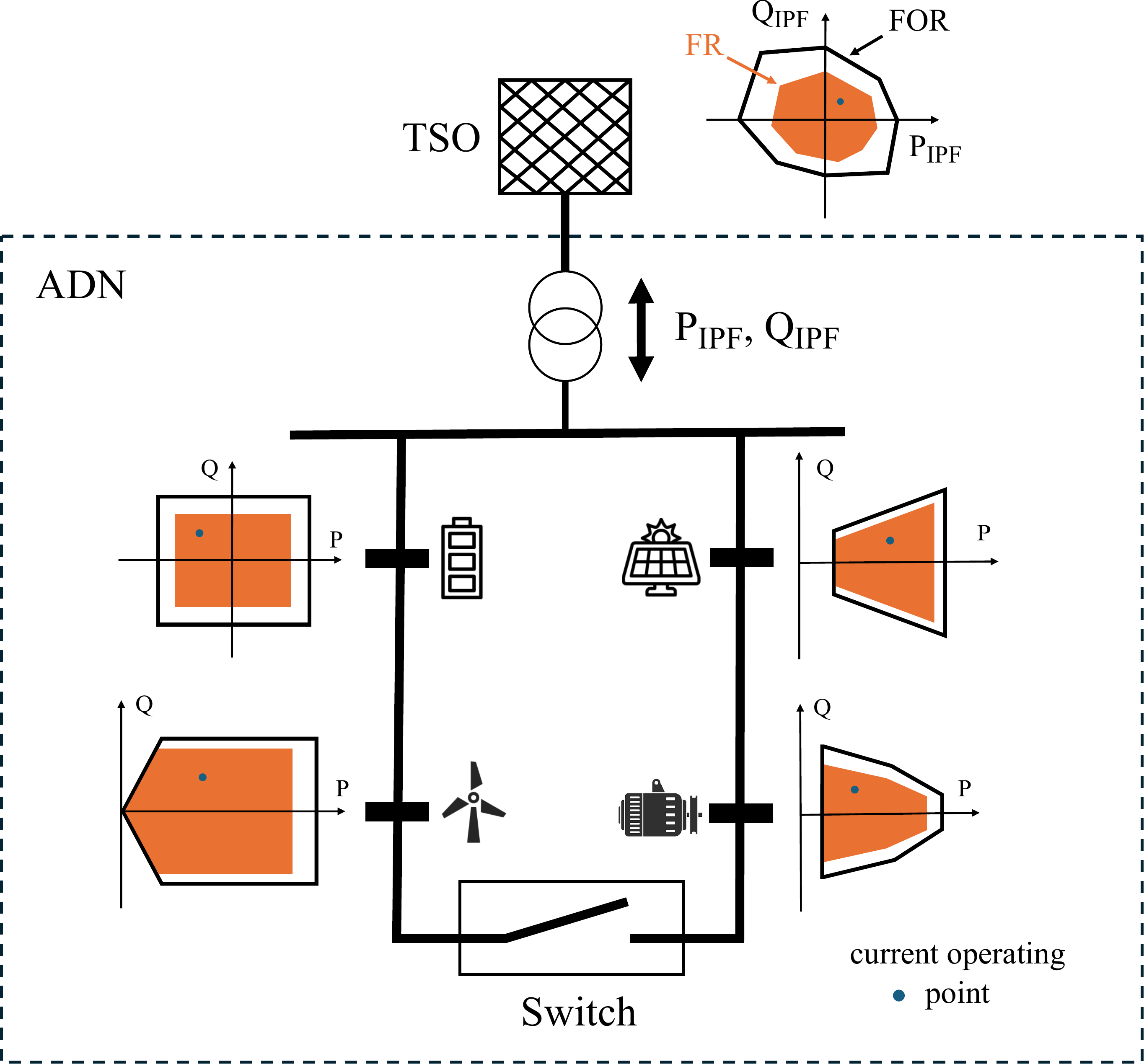}
    \caption{A Network's Feasible Operating Regions and Flexibility Regions}
    \label{Schematic Representation of FORs and FRs}
\end{figure}

Due to relevance for under- and overvoltage incidents~\cite{Cutsem.2008}, recent studies have focused on calculating $P$- and $Q$-flexibility.
Furthermore, various methods have been proposed to obtain FORs including those using computational geometry, random sampling or OPFs.
For example, Minkowski Sums handle generic polygonal shapes to provide FORs as in~\cite{Kundu.2018}.
However, crucial grid constraints are not considered often causing problems.
Applying random sampling is simple but rather time consuming due to the vast number of calculations needed for high-quality results.
One such approach are Monte-Carlo simulations like in~\cite{Heleno.2015}, which define random setpoints for controllable assets before checking solutions for the violation of limitations.
Third, OPF methods have been examined providing techniques for solving the problem efficiently.
Advantages are the reduction of runtime and improved effectiveness.

\section{Determination of Feasible Operating Regions and Flexibility Regions}\label{Determination of FORs and FRs}

\subsection{Setup}\label{Setup}

\begin{figure}[h]
    \centering
    \includegraphics[width=0.35\textwidth]{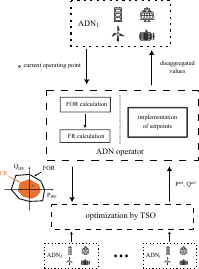}
    \caption{Conceptual Framework of Coordination Processes}
    \label{Conceptual Framework of Coordination}
\end{figure}

Within this paper, we consider different types of renewable energy sources to be connected to a radial network as shown in~\Cref{Schematic Representation of FORs and FRs}.
Each asset can be represented by a characteristic FOR for which a time-dependent FR exists, following~\cite{Contreras.2019b}.
Furthermore, such architecture is able to provide an Interconnected Power Flow (IPF) at the PCC caused by aggregating assets' current setpoints.
Consequentially, Transmission~(TSOs) and Distribution System Operators~(DSOs) must be coordinated to ensure ancillary services being activated securely.
Hence, we propose a conceptual framework where responsibilities are managed upfront as illustrated in~\Cref{Conceptual Framework of Coordination}.

Similarly to~\cite{Prionistis.2023}, we acknowledge the need for an ADN operator.
Initially, it must calculate the FOR once for a specific network topology and pool of assets.
If nothing changes in either aspect, the FOR remains the same throughout upcoming processes.
Following the FOR, FR calculations must be performed with a pre-defined granularity, e.g. once per minute.
The ADN operator must then transmit obtained information to its TSO, where data of other connected ADNs can be summarized.
Depending on needs, the TSO optimizes setpoints of individual ADNs, which again are to be addressed at their PCC as a new IPF like proposed in~\cite{DMG.2018}.
By transmitting~$P^{\text{ref}}$ and~$Q^{\text{ref}}$ back to each ADN operator, the TSO's optimization and control process concludes.
ADN operators are ultimately responsible for breaking target values down to each individual system handled.
In the end, by utilizing assets' flexibility, new operating points are achieved without violating constraints.

\subsection{Non-linear OPF exploration}\label{Non-linear OPF exploration}

Using ac-OPF to calculate FOR boundary points to determine the capabilities of given networks requires formulating the problem with a set of linear/non-linear equality and inequality constraints.
In the course of which we use an objective function \eqref{eq:non-linear objective function} that includes $\alpha, \beta \in \{-1,0,1\}$ as weighing factors like proposed in~\cite{Silva.2018} to define the search direction.
\begin{equation}
    \text{minimize} \ f(x) = \alpha \cdot P_{\text{IPF}}(x) + \beta \cdot Q_{\text{IPF}}(x)
    \label{eq:non-linear objective function}
\end{equation}
The power flow equations and operating constraints are represented by~\eqref{eq:non-linear small p}-\eqref{eq:non-linear voltage boundaries}.
Let $\mathcal{N}$ be the set of buses, $\mathcal{E}$ the set of lines and $\mathcal{G}$ the subset of buses with renewable energy sources connected.
The complex power $s_{i} = p_{i} + j \cdot q_{i}$, which is fed in at each bus $i \in \mathcal{N}$, can then be expressed by
\begin{align}
    p_{i}(V,\theta) = |V_{i}| \sum_{j=1}^{n} y_{ij} |V_{j}| \cos(\theta_{ij} - \delta_{ij}), & \label{eq:non-linear small p} \\
    q_{i}(V,\theta) = |V_{i}| \sum_{j=1}^{n} y_{ij} |V_{j}| \sin(\theta_{ij} - \delta_{ij}), & \label{eq:non-linear small q}
\end{align}
where $n$ is the number of buses.
Both equations depend on complex nodal voltages $V_{i} = |V_{i}| \phase[2]{\theta_{i}}$, voltage angle differences $\theta_{ij}$ and complex admittances $y_{ij} = |y_{ij}| \phase[2]{\delta_{ij}}$ of each line $(i,j) \in \mathcal{E}$.
The nodal power balance in regular steady-state operation is furthermore described by
\begin{align}
    \Delta p_{i} = -p_{d,i} + \sum_{g \in \mathcal{G}} p_{g,i} + p_{i}(V,\theta), & \label{eq:non-linear power balance in p} \\
    \Delta q_{i} = -q_{d,i} + \sum_{g \in \mathcal{G}} q_{g,i} + q_{i}(V,\theta), & \label{eq:non-linear power balance in q}
\end{align}
where $p_{g,i}, q_{g,i}$ represent generation and $p_{d,i}, q_{d,i}$ consumption for each bus $i\in \mathcal{N}$.
These equations show the difference between expected and calculated complex power at bus $i$.

The load flow problem aims to solve non-linear power flow equations, meaning complex bus voltages $V_{i}$ and angles ${\theta_{i}}$ that solve $\Delta p_{i} \approx 0$ and $\Delta q_{i} \approx 0$.
For each line $(i,j) \in \mathcal{E}$,  we apply $V_{i}$ and ${\theta_{i}}$ to
\begin{align}
    P_{ij} = & + g_{ij} \left(|V_{i}|^2 - |V_{i}| |V_{j}| \cos(\theta_{ij})\right)\label{eq:non-linear power flow in p} \\
    & - b_{ij} |V_{i}| |V_{j}| \sin(\theta_{ij}),\notag \\
    Q_{ij} = & - b_{ij} \left(|V_{i}|^2 - |V_{i}| |V_{j}| \cos(\theta_{ij})\right)\label{eq:non-linear power flow in q} \\
    & - g_{ij} |V_{i}| |V_{j}| \sin(\theta_{ij}),\notag
\end{align}
where $g_{ij}, b_{ij}$ are line series admittances and susceptances.
This permits the computation of line flows through all lines.
The IPF, which is the power flow through the slack bus, is included in those equations.
Here, for each line $(i,j) \in \mathcal{E}$, line capacity limits $S_{ij,\text{max}}$ are enforced by
\begin{equation}
    P_{ij}^2 + Q_{ij}^2 \leq S_{ij,\text{max}}^2
    \label{eq:non-linear line capacity limit}
\end{equation}
and voltage boundaries for each bus $i\in \mathcal{N}$ by
\begin{equation}
    V_{i,\text{min}} \leq V_i \leq V_{i,\text{max}}.
    \label{eq:non-linear voltage boundaries}
\end{equation}

\section{Accelerated calculation methods}\label{Accelerated calculation methods}

\subsection{Piecewise linearization}\label{Piecewise linearization}

With the strict PieceWise Linearization~(PWL) as introduced in~\cite{Jiang.2020}, the non-linear power flow equations are discretized into linear segments.
By introducing auxiliary variables, the original non-linear problem is transformed into a mixed-integer linear programming problem.
The linear approximation of the cosine function is realized by dividing it into multiple segments.
In~\eqref{eq:pwl power flow in p}, \eqref{eq:pwl power flow in q}, $v_{i} = |V_{i}|^{2}$ depicts the squared nodal voltage magnitudes. The linear combination of non-linear $\cos(\theta_{ij})$, which is divided by breakpoints $\cos(\theta_{ij,s})$ with $N_{s}$ segments for all $(i,j) \in \mathcal{E}$, is displayed by $\sum_{s=0}^{N_{s}} w_{s} \cos(\theta_{ij,s})$ with coefficients $w_{ij,s} \in [0,1]$ of breakpoints.
\begin{align}
    P_{ij} =& + g_{ij} \left(\frac{v_i}{2}-\frac{v_j}{2}-\sum_{s=0}^{N_s} w_{ij,s} \cos(\theta_{ij,s})+1\right)\label{eq:pwl power flow in p}\\
    & -b_{ij} \sum_{s=0}^{N_s} w_{ij,s} \theta_{ij,s} \notag\\
    Q_{ij} =& - b_{ij} \left(\frac{v_i}{2}-\frac{v_j}{2}-\sum_{s=0}^{N_s} w_{ij,s} \cos(\theta_{ij,s})+1\right)\label{eq:pwl power flow in q}\\
    & -g_{ij} \sum_{s=0}^{N_s} w_{ij,s} \theta_{ij,s} \notag
\end{align}

Furthermore, \eqref{eq:pwl quadratic line capacity limit in p+q}-\eqref{eq:pwl quadratic line capacity in q} represent the piecewise linearized quadratic line capacity limits as in~\cite{Chen.2016} for all $(i,j) \in \mathcal{E}$ with
\begin{align}
    \sqrt{2}S_{ij,\text{max}} &\geq P_{ij} + Q_{ij} \geq -\sqrt{2}S_{ij,\text{max}},\label{eq:pwl quadratic line capacity limit in p+q}\\
    \sqrt{2}S_{ij,\text{max}} &\geq P_{ij} - Q_{ij} \geq -\sqrt{2}S_{ij,\text{max}},\label{eq:pwl quadratic line capacity limit in p-q}\\
    S_{ij,\text{max}} &\geq P_{ij} \geq -S_{ij,\text{max}},\label{eq:pwl quadratic line capacity in p}\\
    S_{ij,\text{max}} &\geq Q_{ij} \geq -S_{ij,\text{max}}.\label{eq:pwl quadratic line capacity in q}
\end{align}

Further information on the introduction of auxiliary variables can be found in~\cite{Jiang.2020}.
In addition to the strict PWL, a relaxed PWL has been implemented according to~\cite{Sun.2022} and is analyzed hereafter.
Compared to strict PWL, the relaxed PWL transforms the mixed-integer linear programming problem into a linear programming problem.
Details are discussed in~\cite{Sun.2022}.

\subsection{The DistFlow model}\label{The DistFlow model}

In the DistFlow model, the non-convexity of the ac-OPF is addressed by convex relaxation.
Since ADNs are mostly radial networks, we ignore phase angles of voltages and currents according to~\cite{Farivar.2013}.
Moreover, relaxing quadratic equality constraints to inequality constraints yields a second order cone program as shown in~\cite{Farivar.2013}.
$P_{jk}$ depicts the power transfer going from $j$ to $k$, whereas the power transfer coming from $i$ to $j$ is described by $P_{ij}$.
The power balance equations at each bus $j \in \mathcal{N}$ after angle relaxation can then be described by
\begin{equation}
    \sum_{k} P_{jk} - \sum_{i} (P_{ij} - r_{ij} l_{ij}) = p_{j},\label{eq:DistFlow power balance in p}
\end{equation}
\begin{equation}
    \sum_{k} Q_{jk} - \sum_{i} (Q_{ij} - x_{ij} l_{ij}) = q_{j},\label{eq:DistFlow power balance in q}
\end{equation}
where $l_{ij} = |I_{ij}|^{2}$ are squared line current magnitudes and $r_{ij}, x_{ij}$ respective line series resistances and reactances describing losses.
In addition, Ohm's law after angle relaxation for each line $(i,j) \in \mathcal{E}$ is described by
\begin{align}
    v_{j} = & v_{i} - 2(r_{ij} P_{ij} + x_{ij} Q_{ij}) + (r_{ij}^2 + x_{ij}^2) l_{ij}.\label{eq:DistFlow ohm's law}
\end{align}
Furthermore, the line flow equation for each line $(i,j) \in \mathcal{E}$ after angle and convex relaxation is described by
\begin{equation}
    \left\lVert\begin{array}{c}
    2P_{ij} \\
    2Q_{ij} \\
    l_{ij} - v_{i}
\end{array}\right\rVert_{2} \leq l_{ij} + v_{i},\label{eq:DistFlow line flow equation}
\end{equation}
where $\left\lVert x \right\rVert_{2}$ is the second-order norm, or Euclidean distance.
Since \eqref{eq:DistFlow line flow equation} depicts a second order constraint, the feasible set is convex.
Furthermore, because the objective function is linear, the DistFlow model represents a conic optimization.

\section{Test System Results}\label{Test System Results}

\begin{figure}[b]
    \begin{subfigure}{.4\linewidth}
        \centering
        \includegraphics[width=3cm]{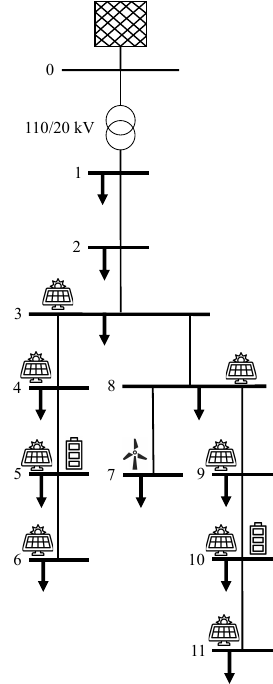}
        \caption{CIGRE MV}
    \end{subfigure}
    \begin{subfigure}{.55\linewidth}
    \centering
    \includegraphics[width=4.7cm]{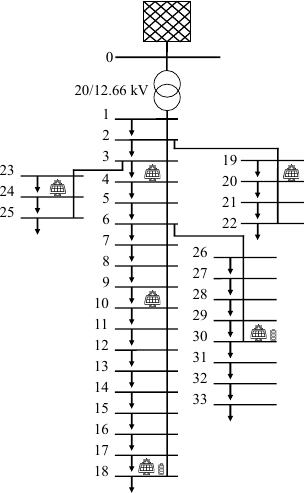}
    \caption{IEEE 33-bus}
    \end{subfigure}
    \caption{Test grids}
    \label{Test grids}
\end{figure}

In this section, we present FORs calculated by the algorithms from~\Cref{Non-linear OPF exploration,Piecewise linearization,The DistFlow model} using two test grids.
The CIGRE benchmark network is taken from pandapower~\cite{Thurner.2018} and adapted with photovoltaic, wind generation and batteries as seen in~\Cref{Test grids}.
Also, the IEEE 33-bus network is taken from MATPOWER~\cite{Zimmerman.2011} and supplied with photovoltaic and batteries accordingly.
All results are compared against each other regarding precision and runtime.
Here, the non-linear OPF is solved using CasADi~\cite{Andersson.2019}.
Furthermore, both PWLs and the DistFlow model are solved using Gurobi Optimizer~\cite{Gurobi.2023}.
Each test is conducted on a PC with an Intel~Core~i5-1135G7 @~2.4~GHz~8~GB~RAM.

\begin{figure}[]
    \centering
    \includegraphics[width=\linewidth]{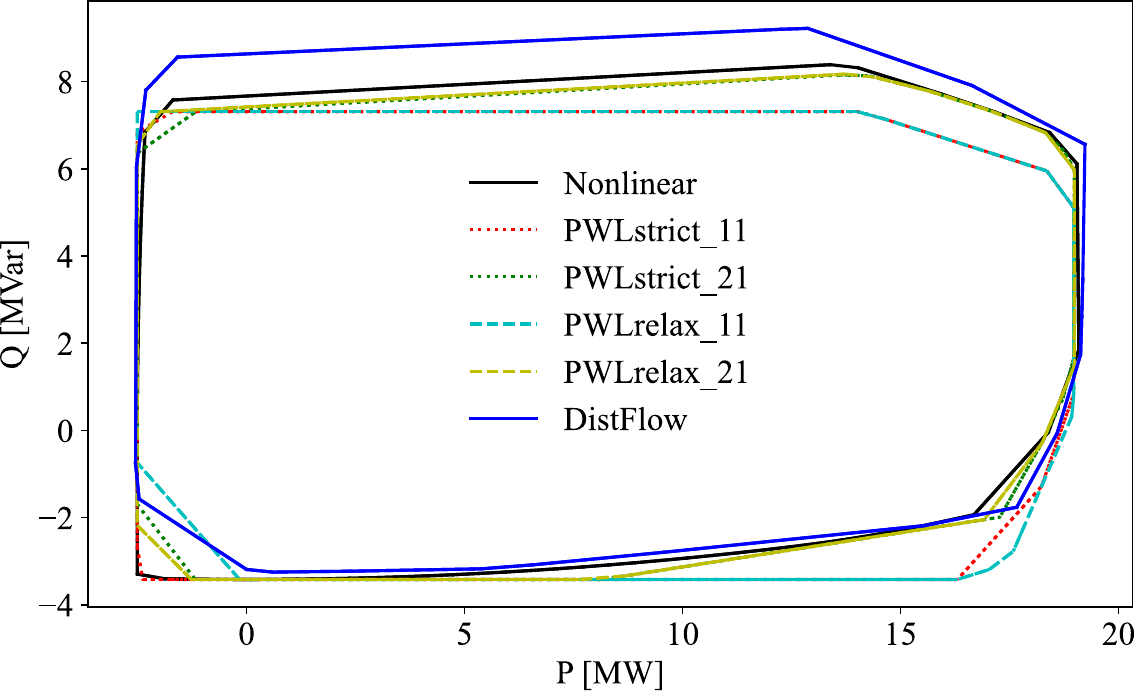}    
    \caption{CIGRE's Feasible Operating Regions using various Algorithms}
    \label{CIGRE-FOR using various algorithms}
\end{figure}

The resulting FORs presented in~\Cref{CIGRE-FOR using various algorithms} are obtained after five iterations.
Each iteration creates new boundary points, that improve the results' accuracy.
Starting with four boundary points initially, the number increases exponentially with the number of iterations.
The algorithms stop as soon as the distance between neighboring boundary points crosses a $5~\%$ threshold compared to the maximum values of $P$ and $Q$ respectively.
Finally, we connect all neighboring points obtaining a polygon, the convex hull of the set, that serves as the FOR.

\begin{figure}[]
    \centering
    \includegraphics[width=\linewidth]{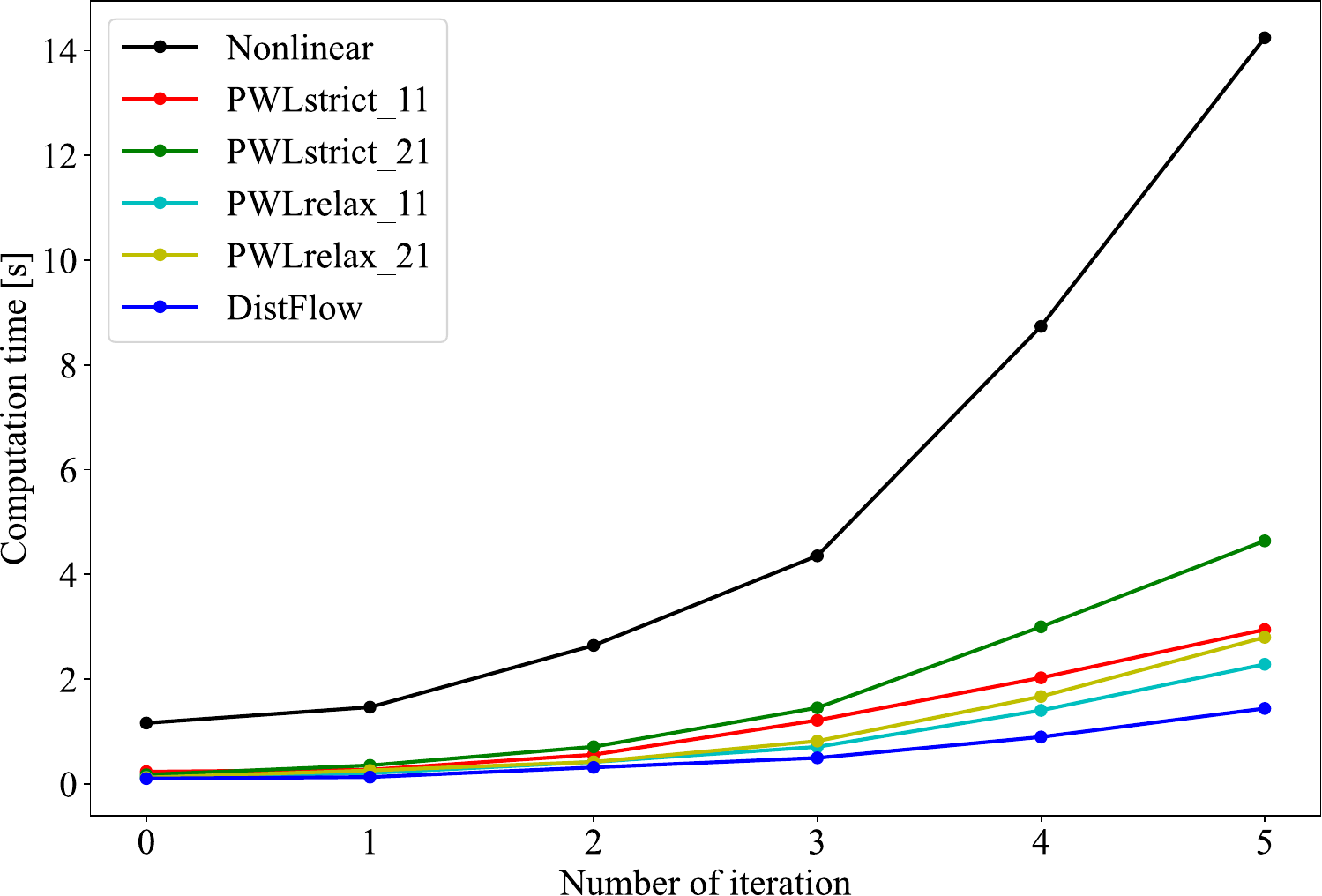}
    \caption{Algorithms' runtimes for CIGRE's Feasible Operating Regions}
    \label{Algorithms' runtimes for CIGRE's FOR}
\end{figure}

We can see that all FORs result in deviations compared to the non-linear case.
Both PWL methods lead to considerably shifted FORs if they contain 11 segments instead of 21.
Interestingly, the DistFlow model leads to errors at higher values for $Q$, while $P$ fits well.
In \Cref{Algorithms' runtimes for CIGRE's FOR}, the algorithms' runtimes are shown.
The results indicate that the DistFlow model is the fastest method examined, followed by relaxed PWL and strict PWL.
Depending on the number of segments, the PWL methods' runtimes change significantly.

\begin{table}[h]
    \caption{Runtimes of CIGRE's and IEEE's Feasible Operating Regions}
    \begin{center}
        \renewcommand{\arraystretch}{1.5}
        \begin{tabular}{|c|c|c|}
            \hline
            \diagbox{\textbf{Algorithm}}{\textbf{ADN's FOR}} &
            \makecell{\textbf{CIGRE [s]} \\ + \textbf{(speed-up)}} &
            \makecell{\textbf{IEEE [s]} \\ + \textbf{(speed-up)}} \\
            \hline
            \textbf{Non-Linear} & 14.24 (-) & 39.98 (-) \\
            \hline
            \textbf{PWLstrict\_11} & 2.95 (4.83) & 7.28 (5.49) \\
            \hline
            \textbf{PWLstrict\_21} & 4.63 (3.08) & 9.09 (4.4) \\
            \hline
            \textbf{PWLrelaxed\_11} & 2.28 (6.25) & 6.78 (5.9) \\
            \hline
            \textbf{PWLrelaxed\_21} & 2.8 (5.09) & 6.72 (5.95) \\
            \hline
            \textbf{DistFlow} & 1.44 (9.89) & 1.78 (22.46) \\
            \hline
        \end{tabular}
        \renewcommand{\arraystretch}{1}
        \label{Execution Times of CIGRE's and IEEE's respective FOR}
    \end{center}
\end{table}

\begin{figure}[h]
    \centering
    \includegraphics[width=\linewidth]{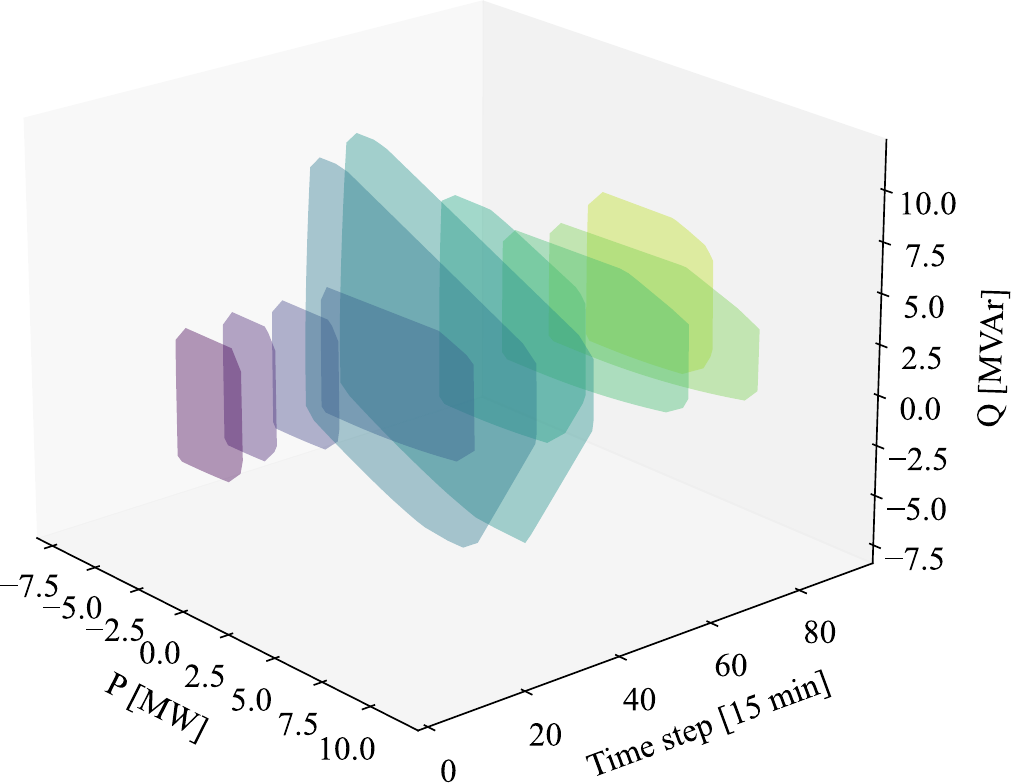}
    \caption{CIGRE's Flexibility Regions from Time Series Data}
    \label{CIGRE Time Series FR}
\end{figure}

In \Cref{Algorithms' runtimes for CIGRE's FOR}, the acceleration obtained by all models stands out compared to the non-linear case. In specific, \Cref{Execution Times of CIGRE's and IEEE's respective FOR} shows the substantial time saving capabilities of the DistFlow model.
Concerning the calculation of FRs, exemplary results are shown in \Cref{CIGRE Time Series FR}, obtained by using time series data from~\cite{Thurner.2018}.
Here, a single day was split into 96 intervals of 15 minutes each, displaying 10 exemplary FRs.
The increasing output of photovoltaic systems during the middle of the day can be seen.

\section{Conclusion and Outlook}\label{Conclusion and Outlook}

In this paper, the applicability of various approaches to calculate FORs/FRs within an existing adaptive control system was discussed. The goal was to improve grid stability by providing ancillary services.
For this purpose, several calculation methods, from linearization to relaxation, were compared against each other.
The simulation results show that the calculation of FORs/FRs to be used in a CPFC is possible within a reasonable time frame.
Therefore, it would be advantageous if these methods were utilized more frequently.
The performance of ADNs thereby essentially depends on available resources, while the computation methods of FORs/FRs can be accelerated significantly by linearization and relaxation.

In future work, the increasing complexity of greater networks should be thematized, accompanied by the challenges of communication.
In addition, laboratory experiments and field tests are planned to extend the analysis proposed in this paper.
In particular, laboratory tests based on Power-Hardware-in-the-Loop will be conducted to characterize the actual FORs/FRs on real components under different scenarios.

\section*{Acknowledgment}\label{Acknowledgement}

Special thanks to Prof. Dr. Costas Vournas from the National Technical University of Athens for valuable discussions.

\renewcommand*{\bibfont}{\scriptsize}

\printbibliography

\end{document}